\documentclass[a4paper,11pt]{article}

\usepackage{amssymb}
\usepackage{amsmath}
\usepackage[belowskip=-8pt]{caption}
\usepackage[round,authoryear]{natbib}
\usepackage{authblk}
\usepackage{colortbl}
\definecolor{webgreen}{rgb}{0,0.4,0}
\definecolor{webbrown}{rgb}{0.6,0,0}
\definecolor{purple}{rgb}{0.5,0,0.25}
\definecolor{darkblue}{rgb}{0,0,0.7}
\definecolor{darkred}{rgb}{0.7,0,0}
\usepackage{hyperref}
\hypersetup{colorlinks,citecolor=darkblue,filecolor=black,linkcolor=darkred,urlcolor=webgreen,pdfpagemode=None,
pdfstartview=FitH}

\newcommand{\ignore}[1]{}

\usepackage{cleveref}
\crefname{claim}{claim}{claims}
\crefname{fact}{fact}{facts}
\crefname{algorithm}{algorithm}{algorithms}
\crefname{observation}{observation}{observations}
\crefname{equation}{equation}{equations}
\crefname{assumption}{assumption}{assumptions}
\crefname{lemma}{lemma}{lemmata}
\crefname{corollary}{corollary}{corollaries}
\crefname{figure}{fig.}{figs.}

\usepackage{pgf}
\usepackage{verbatim}
\usepackage{enumerate}
\usepackage{amssymb}
\usepackage{mathrsfs}
\usepackage{algorithm}
\usepackage[noend]{algorithmic}
\usepackage{mathtools}

\usepackage{resizegather}

\newif\ifverbose
\verbosetrue 

\usepackage{url}

\usepackage{xspace}

\usepackage{nicefrac}

\newcommand{\longversion}[1]{}

\newcommand{\truthbot}{{\bf \texttt{TruthBot}}}
\newcommand{\FCR}{{\bf \texttt{FCR}}}

\usepackage[a4paper, total={6in, 9in}]{geometry}
\usepackage{cleveref}
\crefname{claim}{claim}{claims}
\crefname{fact}{fact}{facts}
\crefname{algorithm}{algorithm}{algorithms}
\crefname{observation}{observation}{observations}
\crefname{equation}{equation}{equations}
\crefname{assumption}{assumption}{assumptions}
\crefname{hypothesis}{hypothesis}{hypotheses}
\usepackage{enumerate}
\usepackage{subcaption}
\usepackage[noend]{algorithmic}
\usepackage{resizegather}
\usepackage{enumitem}
\usepackage{multirow,bigdelim}
\usepackage{cancel}
\usepackage{framed}
\usepackage{wrapfig}
\usepackage{tfrupee}
\usepackage{bbold}
\allowdisplaybreaks

\title{\bf \truthbot: An Automated Conversational Tool for Intent Learning, Curated Information Presenting, and Fake News Alerting}

\author[1]{Ankur Gupta$^*$}
\author[1]{Yash Varun$^*$}
\author[1]{Prarthana Das$^*$}
\author[1]{Nithya Muttineni$^*$}
\author[1]{Parth Srivastava$^*$}
\author[1]{Hamim Zafar}
\author[2]{Tanmoy Chakraborty}
\author[1]{Swaprava Nath}

\affil[1]{\small Indian Institute of Technology Kanpur, $^*$equal contribution, \texttt{\{ankugupt,yashyv,prdas,mnithya,parthsri,hamim,swaprava\}@iitk.ac.in}}
\affil[2]{\small Indian Institute of Information Technology Delhi, \texttt{tanmoy@iiitd.ac.in}}

\date{}

\begin{document}
\maketitle

\begin{abstract}
\noindent
 We present \truthbot\ (name is anonymized), an all-in-one multilingual conversational chatbot designed for seeking truth (trustworthy and verified information) on specific topics. It helps users to obtain information specific to certain topics, fact-check information, and get recent news. The chatbot learns the intent of a query by training a deep neural network from the data of the previous intents and responds appropriately when it classifies the intent in one of the classes above. Each class is implemented as a separate module which   uses either its own curated knowledge-base or searches the web to obtain the correct information. The topic of the chatbot is currently set to COVID-19. However, the bot can be easily customized to any topic-specific responses. Our experimental results show that each module performs significantly better than its closest competitor, which is verified both quantitatively and through several user-based surveys in multiple languages. \truthbot\ has been deployed in June 2020 and is currently running. 
\end{abstract}

\section{Introduction}

Social media has endowed us with a massive volume of information, which has both useful, not-so-useful, and misleading content \citep{vraga2019news,shu2017fake}. 
This is worrisome since a usual netizen's social media time is significant. According to eMarketer\footnote{\url{www.emarketer.com}}, US adults use social media for an average of 12 hours  per day.
During the crisis situations like a global pandemic, people desperately look for solutions and end up consuming unverified information from various online resources including social media.

It is observed that the users often trust the messages they see on social media and instant messengers. A study by \citet{bridgman2020causes} shows that those who receive most of the news from social media are more likely to believe falsehoods about COVID-19. It is also observed that the social media has taken over the role of a news platform and now ranks just behind television \citep{kohut2010americans} which may mask true information for certain users. To help such users reach to the correct information, technology needs to be developed within the social media and instant messengers itself. 

On the other hand, popularity of chatbots among the Internet users has increased significantly in the past few years\footnote{https://g.co/trends/7tkFb} \citep{GrandViewResearch2017}. Studies on US-based users have shown that 86\% of them prefer chatting with a chatbot than a human agent \citep{Forbes.com2019}. A study by \citet{brandtzaeg2017people} shows that the major motivations for using chatbots are their timely and efficient assistance for information which improves productivity.
Hence, the prospect of a chatbot to truth-seek information is bright, and serves as the motivation of this paper.

\subsubsection{State-of-the-art and limitations.} The current fact-checking apps require the users to enter the suspicious messages in those apps to fact-check or ask the users to read the current fact-checking articles to find out if their answers lie within the articles. Such apps are not very {\em useful}, since  suspicious messages typically come via social media platforms or instant messengers and the cost of switching apps or read verified messages to find out the truth is quite high. The other kind of solutions that involve forwarding service (e.g., to WhatsApp business accounts) for fact-checking are typically manually verified and responded by a team of journalists. The chatbot approach for fact-checking is very limited as we discuss in \Cref{sec:literature}. However, those approaches also do {\em not} (a)~consider a complete {\em truth-seeking} design that provides holistic information on a topic, rather take care of certain frequently asked questions on a topic (e.g., COVID-19), and (b)~handle low-resource languages.

Note that we consider the term {\em truth-seek} to be more general than {\em fact-check}. While fact-checking tries to classify a piece of news to be true or not, truth-seeking provides a complete information against a query. For instance, the query can be a general question about a topic or a news article about which a user is only partially aware or an area-based infection statistics of a disease. The scope of truth-seeking, therefore, subsumes fact-checking as we discuss in the following section.

\subsection{Proposed Solution: \truthbot.} We introduce \truthbot, a truth-seeking chatbot that  (a)~provides answers to topic-specific frequently asked questions (FAQ), (b)~returns articles  after searching on the web, (c)~responses to custom queries (e.g., area-wise infection statistics for COVID-19), in addition to (d)~fact-checking news articles. 
  \truthbot\ uses a deep neural network to identify the intent of the user's query. However, if the confidence of the identification is not significant, it consults with the user through a conversation to make sure that the right intent is conveyed to the chatbot. 
  If the query topic falls in any of the four classes discussed above, it triggers the corresponding module and retrieves the answer. The modules have dependencies, e.g., if the FAQ module cannot respond to the query, it automatically searches for the query in the fact-check or Google search modules. The details are in \Cref{sec:arch}.

\subsection{Evaluation.} We evaluate \truthbot\ in three aspects:
\begin{itemize}
    \item We measure the {\em response accuracy} by finding the relevance of the content of the response with the content of the original query.
    \item We conduct a survey on the {\em satisfaction} of the chatbot regarding the topic it handles (COVID-19 in our example).
    \item We also conduct another survey with a different population on the {\em user interface} about the usefulness, ease of use, credibility, precision, and value of the chatbot.
\end{itemize}
\Cref{tab:comparison} shows a typical example of the responses returned by the {\em international fact-checking network (IFCN)} chatbot on WhatsApp\footnote{https://www.poynter.org/fact-checking/2020/power-in-your-pocket-how-to-use-the-ifcns-new-whatsapp-chatbot/} vis-a-vis \truthbot, where the latter is much more precise against the query.
\begin{table}[t]
    \centering
    \begin{tabular}{p{\linewidth}}
    \hline
    {\bf Query:} monkeys are being tortured in laboratories to find COVID-19 vaccine \\ \hline \hline
         {\bf IFCN chatbot} (click \url{https://wa.me/17272912606?text=hi})  \\\hline
         {\bf Response:} (1) Claim rated False: A vaccine for coronavirus was available since 2001. \\
         Fact checked by Newsmeter.in on 04/26/2020\\
         This is a dog vaccine wrongly claimed as being a vaccine for COVID-19. (provides the link) \\
        (2) Claim rated False: A WhatsApp forward claims that the COVID-19 vaccine is ready, was announced by U.S. President Donald Trump, and will be launched next Sunday ... \\
        Fact checked by Newschecker on 03/23/2020 (provides the link)\\ \hline \hline
         {\truthbot} \\\hline
         {\bf Response:} (1) (provides link to news which reads) \\
         Fact check: This disturbing viral video of animal cruelty is not of COVID-19 vaccine test \\
         A disturbing video purportedly showing animal cruelty circulating on social media claims: this is how monkeys are being tortured in laboratories to test the COVID-19 vaccine.\\
         Viral video showing animal torture is from last year reportedly from a German laboratory ...\\
         (2) (provides link to news which reads) \\
         Ferrets are among the animals researchers are infecting with the Covid-19 virus as they search for a suitable species to use in tests ... \\\hline
    \end{tabular}
    \caption{An example query and response obtained from two chatbots  --  IFCN (a baseline) and \truthbot\ (our method).}
    \label{tab:comparison}
\end{table}

\subsection{Our contributions.} The contributions of this paper can be summarized as follows:
\begin{itemize}
    \item Unlike other fact-checking chatbots, \truthbot\ is a truth-seeking bot, which performs  four tasks (\Cref{sec:faq} to \Cref{sec:google}) via a conversational AI engine (\Cref{sec:conversation}) which identifies the intent of a query and guides the user through a piece of curated information or gets the intent clarified from the user. 
    \item The chatbot identifies the query language and responds in that language (\Cref{sec:multilingual}). This is supported for the 108 languages that Google translate is capable of handling (we use the Google translate service for this task), which is a major advantage of using \truthbot, particularly for low-resource languages.
    \item Experiments (\Cref{sec:experiments}) show that the response accuracy for all types of queries is much better than that of the IFCN chatbot, which also does a query-based fact-checking and is the closest in its design to our chatbot.
    \item \truthbot\ has been operational since June 2020 and, it is currently accepting queries related to COVID-19.
\end{itemize}

\subsection{Reproducibility} We provide all data, code, and survey results separately.

\section{Related work}
\label{sec:literature}
We discuss the related studies briefly in two aspects -- fake news detection and task-specific chatbots.

Detecting malicious content on social media such as fake news, misinformation, disinformation, rumors, cyberbullying has been one of the primary research agenda in natural language processing and social computing. Due to the abundance of literature, we would like the readers to point to two recent survey articles by \citet{zhou2018survey} and \citet{bondielli2019survey}. Existing approaches can be broadly divided into four categories -- (i) {\em knowledge-based}, which verifies whether the given news is consistent with a knowledge base (true knowledge) \cite{pan2018content}; (ii) {\em style-based}, which employs various stylometry analysis modules to check the similarity of the writing style of a given news with existing fake news \cite{zhou2019fake}; (iii) {\em propagation-based}, which explores the propagation of a given news on social media to check how it correlates with the usual fake information diffusion \cite{bian2020rumor}; and (iv) {\em source-based}, which examines the trustworthiness of the source (the account) from which the news has started propagating \cite{zhou2019network}.


Another direction of research relevant to the current study deals with the use of chatbots as an assistant for different applications \citep{lin2020caire,radziwill2017evaluating,folstad2018chatbots}. Though chatbots originated in the 1960s, its use as an AI assistant and widespread acceptance is a recent phenomenon \citep{shum2018eliza}. The use of chatbots for fake news detection is quite a recent development. The approach closest to this paper is by the {\em international fact-checking network (IFCN)} \citep{IFCN2020}, that has tied up with WhatsApp to develop a fact-checking chatbot on that platform for people to fact-check news articles. In this paper, we provide case-studies and a comparison of outputs of various queries of our chatbot vis-a-vis the IFCN chatbot.

\truthbot\ is quite different from all chatbot approaches to true information retrieval, since it is not limited to fact-checking alone, rather provides a complete information on a topic in multiple languages.

\section{Architecture of \truthbot}
\label{sec:arch}

\begin{figure}
    \centering
    \includegraphics[width=\linewidth]{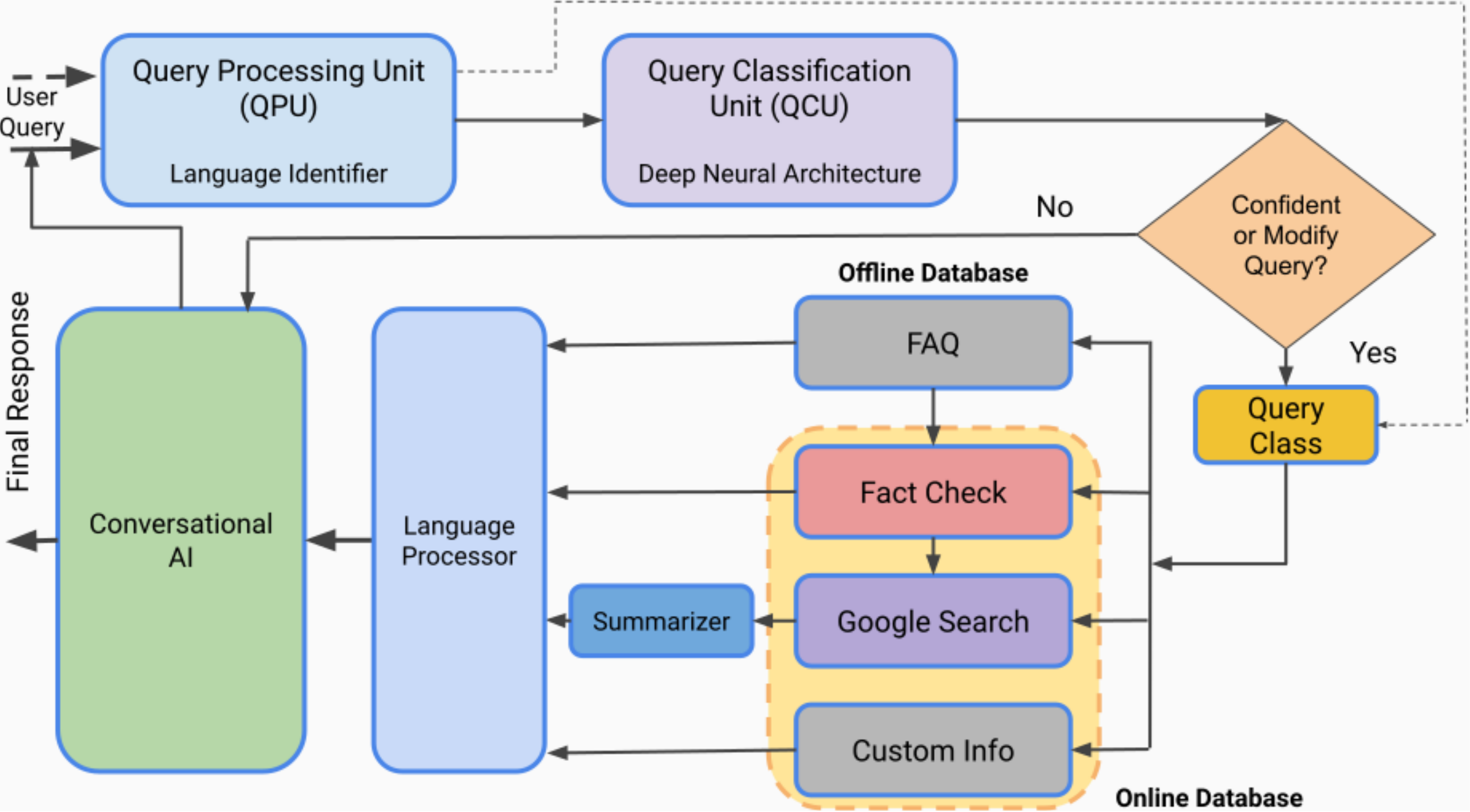}
    \caption{Schematic workflow of \truthbot.}
    \label{fig:architecture}
\end{figure}

The objective of \truthbot\ is to bring all relevant truth-seeking activities on a topic within the scope of a single chatbot. Users can interact with \truthbot\ to get (a)~general information regarding the topic, (b)~fact-check potentially fake news, (c)~general information on search queries from Google search, and (d)~response from other custom information modules. The current setup of \truthbot\ is tuned to provide all such information regarding COVID-19 pandemic, with the custom information module being the infection and death statistics of a city, district, state, or country. However, it is perfectly general to be tuned to any other topic with a little customization.
We present the basic architecture of \truthbot\ in this section. \Cref{fig:architecture} shows the graphical outline of the working units of \truthbot.

\subsection{User queries}
\label{sec:dataset}

The beta-version of the bot has been operational since June, 2020, and has attracted 696 unique users (who either have directly used or are informed about the bot till August, 2020) on the three platforms -- WhatsApp, Facebook Messenger, and Telegram. For the analyses presented in this paper, we used about 1300 queries which were received till July, 2020.\footnote{Collectively we got 2270 different queries sent to our bot till September 9, 2020. The performance of the bot will improve as this training set becomes richer.} We manually classified these queries into {\em six} classes -- {\em frequently asked questions} (FAQ) on a specialized topic (COVID-19 in our case), {\em fake information verification} (FAKE), {\em general search for a news} (GEN), {\em some custom information} (area-wise infection, death statistics--AREASTAT--in our case of COVID-19), {\em greeting}, and {\em spam}. Some of the queries, e.g., ``{\em warm saline water gargling can cure coronavirus}", were classified into multiple classes, as this example can be classified as FAQ, FAKE, and GEN. This manual labeling task took about 30 person-hours since one needed to read every query and use human justification to classify them. This process generated the training set (the JSON file with these classification is available in the supplementary material) which we used to classify new queries.

\subsection{Conversational AI approach}
\label{sec:conversation}

\begin{figure}[t]
    \centering
    \includegraphics[width=\linewidth]{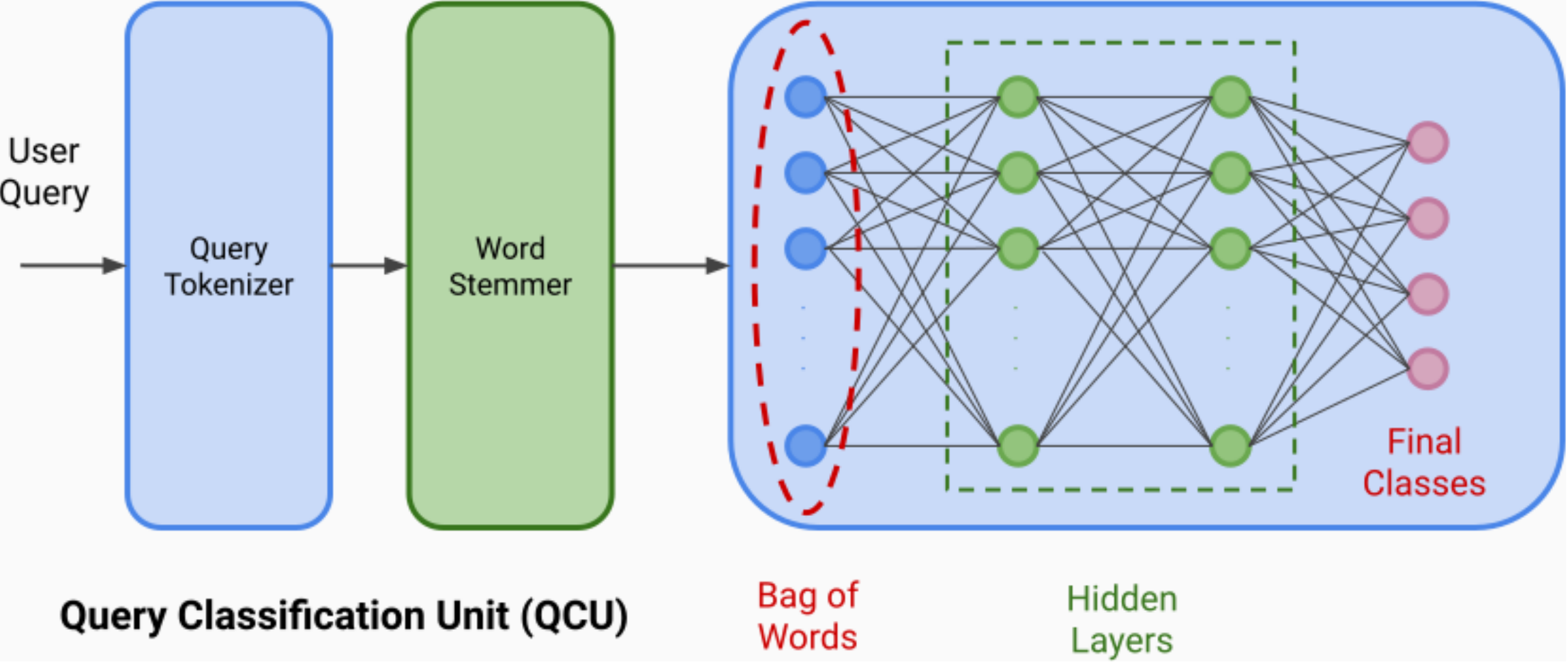}
    \caption{Deep neural network for query classification (number of nodes is representative).}
    \label{fig:qcu}
\end{figure}

\begin{figure}[h!]
    \centering
    \includegraphics[width=0.9\linewidth]{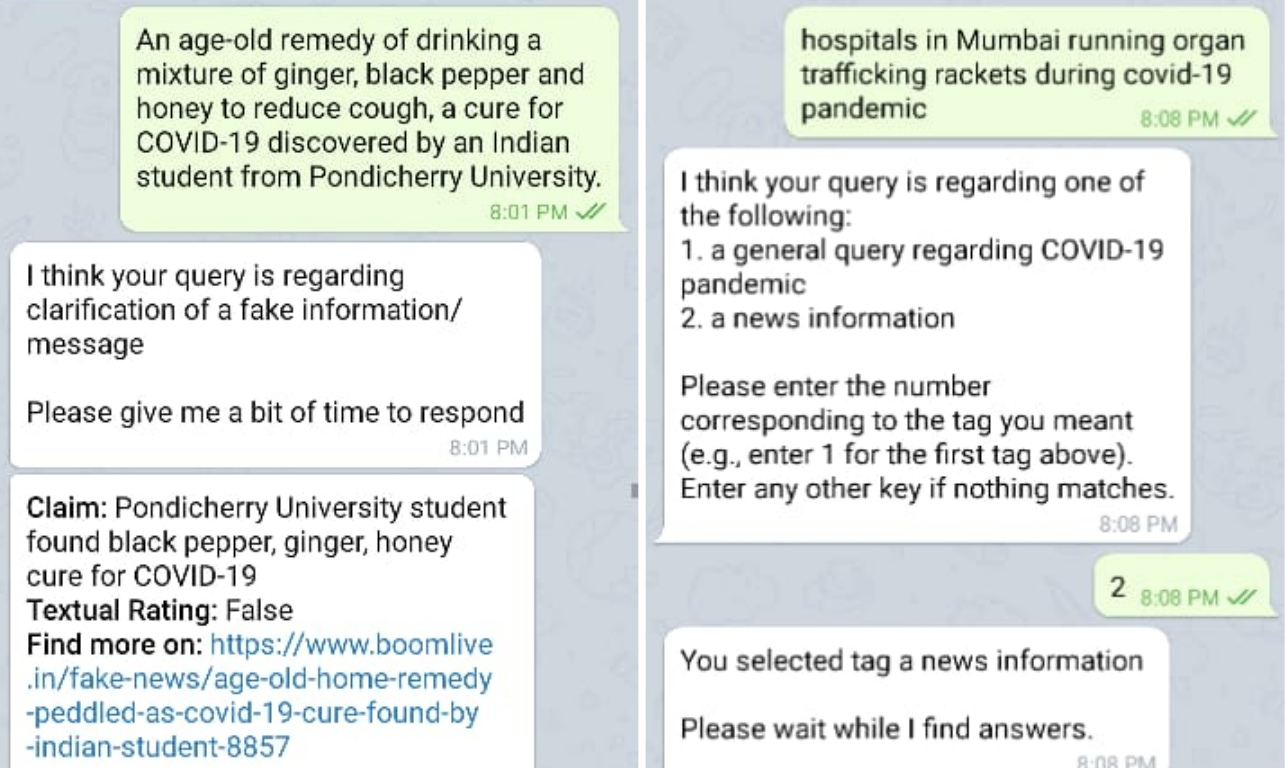}
    \caption{Examples of the conversation with the user when \truthbot\ is confident (left) and not completely confident (right).}
    \label{fig:conv_ai}
\end{figure}

\truthbot\ is designed to help users check the truth of a news or a message. The focus is to provide the least cognitive load to the user and understand the intent of her queries before proceeding on to searching the information. This is important since the trajectory of the search for the different tasks \truthbot\ is capable of doing is quite different. The query is tokenized and stemmed into root words to create a bag of words. The collection of such words and the query classes are used to train the Query Classification Unit (QCU). This unit classifies the intents of the queries into the six classes discussed in \Cref{sec:dataset} using a {\em deep neural network} (DNN). Any new query is soft-classified into these classes using the softmax scores. If the scores are significant---the threshold has been tuned based on manual inspection of the classification---then the action corresponding to that class is invoked (e.g., it calls the module Fact Check if it is classified as a fake information verification query). If the scores of classification are not significant, then it returns to the user and provides a selection choice of the intent to make a focused truth-search of the query. A couple of examples of how the conversation unfolds in this method are provided in \Cref{fig:conv_ai}. \Cref{fig:qcu} shows the schematic representation of QCU. The performance metrics of this classification for different classes are shown in \Cref{tab:classification}.
\begin{table}[H]
    \centering
    \begin{tabular}{c||c|c|c}
         Metric~\citep{shu2017fake} & FAQ & FAKE & GEN \\ \hline\hline
         Precision 	& 0.44 & 0.22 & 0.58 \\
         Recall & 0.61 & 0.08 & 0.64 \\
         F1 score & 0.51 & 0.12 & 0.61 \\
         Accuracy & 0.67 & 0.602 &  0.68 \\ \hline
    \end{tabular}
    \caption{Performance metrics of the DNN classification.}
    \label{tab:classification}
\end{table}

\subsection{Topic-specific frequently asked questions (FAQs)}
\label{sec:faq}

The purpose of this module is to answer some standard questions users may have on the topic \truthbot\ is tuned to. Since standard answers to these FAQs are already available, it is possible to create a knowledge-base from which these questions can be directly answered and reduce the latency of the responses. In the following subsection, we discuss the case study of the FAQs on COVID-19 that this bot is currently tuned to.

\subsubsection{Case study: COVID-19.}
\label{sec:faq-covid}

\begin{figure}
    \centering
    \includegraphics[width=\linewidth]{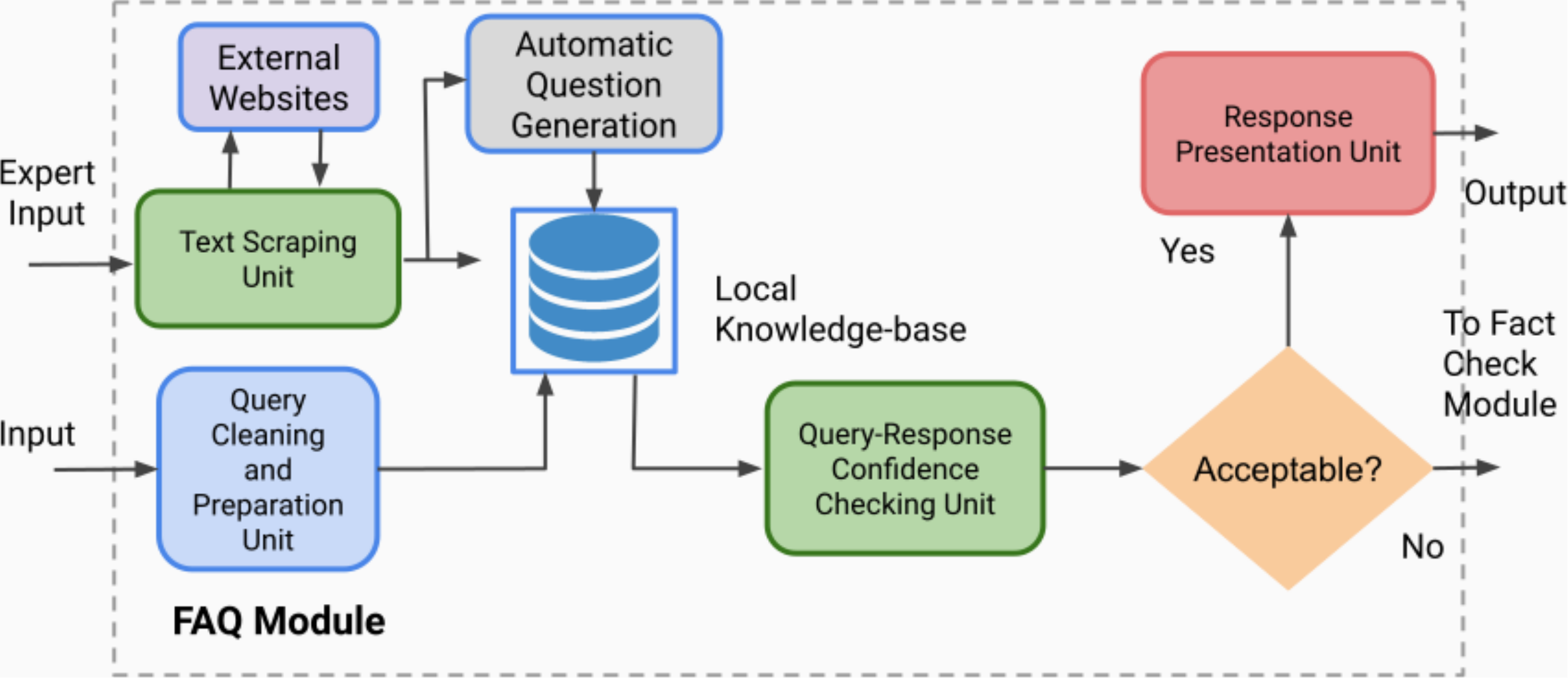}
    \caption{Architecture of the FAQ layer.}
    \label{fig:faq}
    \vspace{-5mm}
\end{figure}

For COVID-19, websites like WHO (\url{https://covid19.who.int/}), CDC (\url{https://www.cdc.gov/}), several government health departments (e.g., the Indian Ministry of Health and Family Welfare (\url{https://www.mohfw.gov.in/})) provide a large collection of FAQs and myth-busters. We first create a knowledge-base by {\em web-scraping} those articles from these websites. \Cref{fig:faq} shows the flowchart of this module.

The Text Scraping Unit (TSU) consists of several \textsf{Python} libraries. The TSU downloads the HTML content of the target webpage (e.g., the FAQ page of CDC\footnote{\url{https://www.cdc.gov/coronavirus/2019-ncov/faq.html}}) by sending an HTTP request to the associated URL and uses it to construct a parse tree using  {\tt html.parser}. The useful FAQs and the paragraphs answering each FAQ are extracted from the parse tree and stored in the local knowledge-base.


To augment the local knowledge-base, we also employ automatic question generation from the answer texts. This is done such that a slightly rephrased question gets a better similarity with one of the questions in our knowledge-base. The answer, however, remains same for all the questions generated from that text. The most straightforward means for this task is answer-aware question generation \cite{lopez2020transformerbased}. In answer-aware question generation, the model is presented with the answer and a passage to generate a question for that answer considering the passage as the context. As the answer-aware models require answers for generating questions, we first split the passage into sentences and select a subset based on named-entity recognition and noun-phrase extraction. We have used a text-to-text transformer model \cite{raffel2019exploring} for question generation. The model is fine-tuned in a multitask way using task prefixes with the help of SQuADv1 dataset \cite{rajpurkar2016squad} along with pre-trained transformers (specifically seq-2-seq models) on paragraphs that provide answers to COVID-19 related FAQs extracted from online databases.

When a user sends the query, we need to find the appropriate question that matches  this query. We use BERT~\cite{devlin2018bert} fine-tuned on CORD-19 dataset\footnote{\url{https://allenai.org/data/cord-19}} to generate the contextual embedding for each {\em question} sentence in the corpus. Then, we use cosine similarity to match a query with one of the questions in the local knowledge-base. A cosine similarity score of $0.85$ is used as an empirically determined threshold for the correctness of the matchings. If a query scores above this threshold, we return the answer corresponding to the matched question in the corpus. Otherwise, the query is transferred to Fact Check module. 

\subsection{Custom information query}
\label{sec:custom}
Several topic-specific custom modules can be added to \truthbot. The purpose of these modules varies with the kind of topic we tune the chatbot to. As an example, we explain one such custom information module for COVID-19 below.

\subsubsection{Case study: COVID-19.}
\label{sec:custom-covid}
Currently, \truthbot\ is running a custom information module, called AREASTAT. The goal of this module is to provide the user with the infection, death, recovery statistics of a city, state, or country. Currently, the bot responds to the finer AREASTAT queries on Indian states and cities. For other countries, it provides country-level aggregate information. However, this can be easily extended with an appropriate database providing the finer information for the cities and states of other countries. 

After the intent of the query is classified as AREASTAT, it queries the databases with the name of the city, state, or country extracted from the query. The following workflow is repeated at most twice until the result is found.
\begin{itemize}
    \item The name of the place is extracted from the query.
    \item The place is looked up in three databases -- state wise, district wise, and country wise in order.
    \item If a match is found in the database, the results of the number of COVID-19 cases, partitioned into {\em confirmed, recovered, active}, and {\em deceased} are sent to the user.
    \item If no match is found in the database, the user is sent a message asking to enter only the name of the place to ensure that there is no spelling mistake.
\end{itemize}
\begin{figure}
    \centering
    \includegraphics[width=\linewidth]{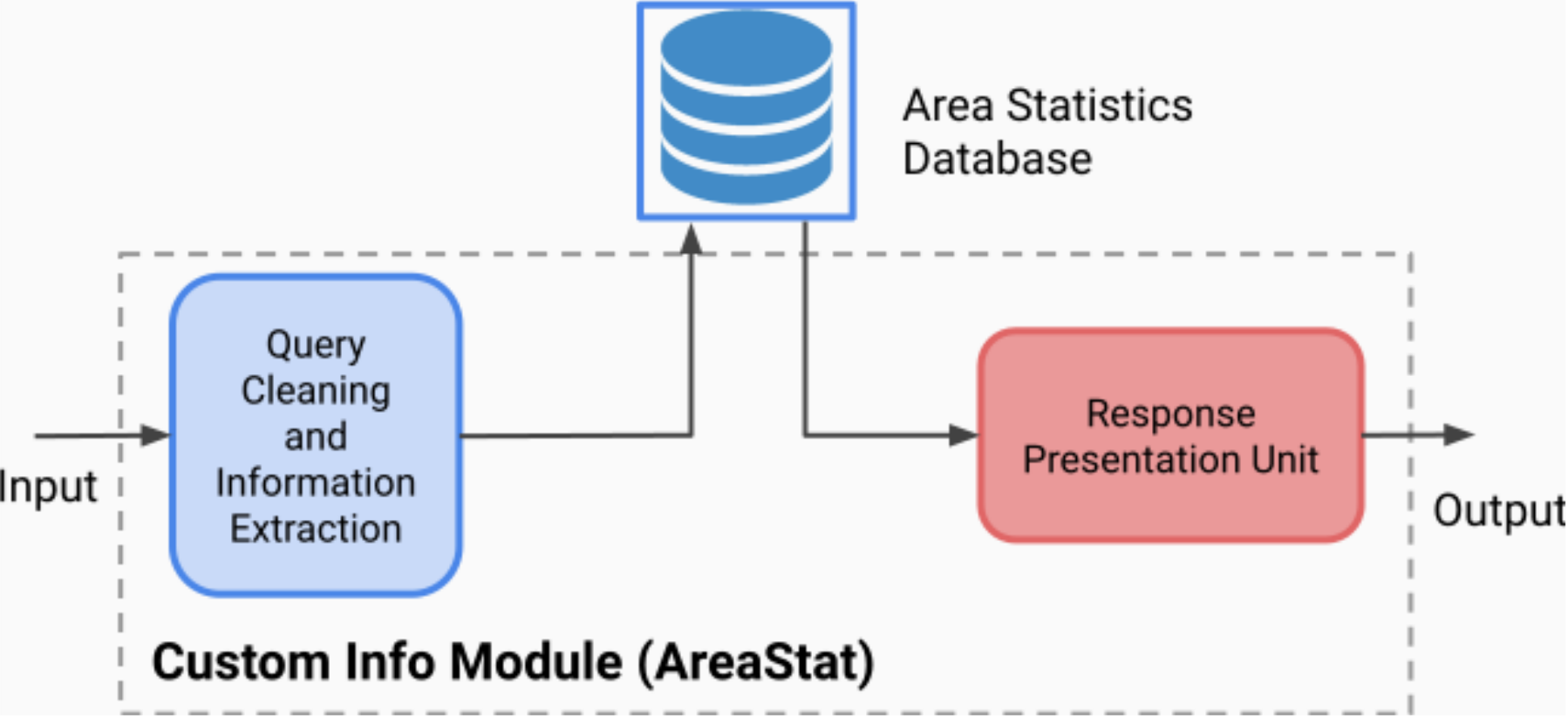}
    \caption{Custom information query (AREASTAT) layer.}
    \label{fig:areastat}
    \vspace{-5mm}
\end{figure}

\subsubsection{Verifying if the query is about AREASTAT.}
The AREASTAT queries follow a pattern, e.g., ``what are the number of cases in $\langle$place name$\rangle$'', ``number of COVID-19 cases in $\langle$place name$\rangle$", ``how many people are infected with covid in $\langle$place name$\rangle$" etc. The query is first processed for the common phrases such as ``number of cases" or ``how many'' followed by the place name. 

However, since the database for area-wise statistics available to us is only for COVID-19, we need to distinguish similar queries that ask the statistics for a different topic, e.g., the death in a cyclone. We first identify if the query has any COVID-19 related keywords along with the phrases discussed above.
Once the query is confirmed to be a COVID-19 statistics query, we detect the place name using the usual patterns of such AREASTAT questions. 

There are two ways in which the statistics query may not be responded to: (a) if the statistics asked is not related to COVID-19, or (b) if the place name is not available in the database. In the former case, the query is transferred to the GEN module for a Google search of the query. In the latter, the user is responded with a message asking to re-enter the place correctly.

\smallskip \noindent
\textit{Statistics databases.}  
The covid19india API\footnote{\url{https://api.covid19india.org/}} crowdsources the latest number of COVID-19 cases from various sources pertaining to each state and city of India. 
%
For the countries other than India, the data is scraped from the Worldometers website\footnote{\url{https://www.worldometers.info/coronavirus/}} at the time when the query is received.



\subsection{Fake news alerting}
\label{sec:fake}

\begin{figure}
    \centering
    \includegraphics[width=\linewidth]{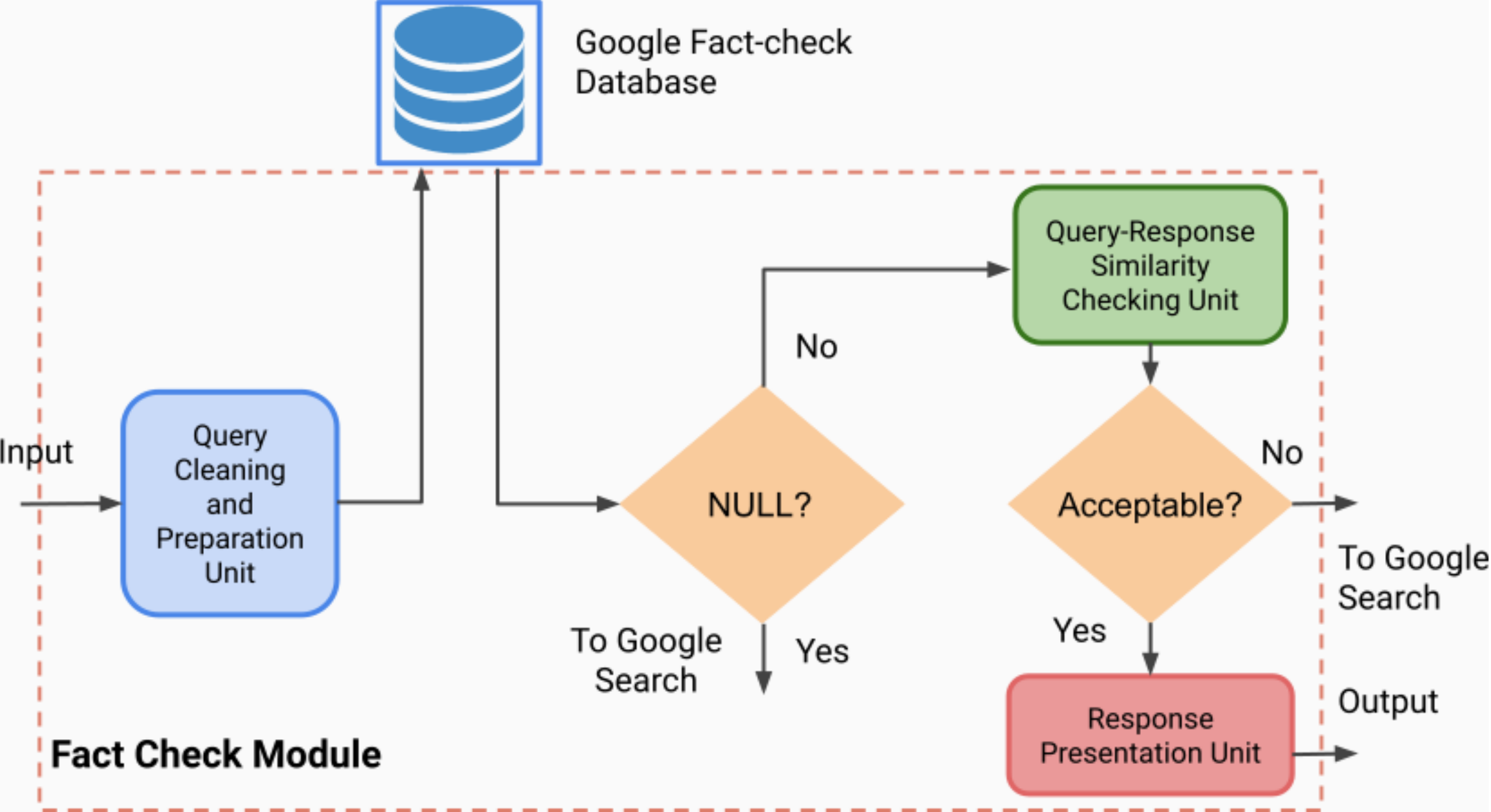}
    \caption{Architecture of Fact Check module.}
    \label{fig:fake}
\end{figure}

Several responsible news agencies bust fake news and post such analysis on their dedicated fact-check section of their websites. If a query is classified as FAKE, it is processed through the Fact Check module that uses `Google Fact Check Claim Search API' to search for the already fact-checked claims by sending an HTTP request. A typical url for calling the API consists of the query, language code, page size and an API key. An API call can return with a nested dictionary (denoted by \FCR) of multiple objects corresponding to the number of fact-checked claims available on different fact-checker websites (e.g., Alt News\footnote{\url{https://www.altnews.in/}}). For each of the retrieved results, we store the TEXT, TEXTUAL RATING, and URL attributes. For each result $\FCR_{i}$, we assign a relevance score $rs_i$ that computes the semantically matching similarity (based on cosine similarity) between the contextual embedding of the TEXT attribute of $\FCR_{i}$ and the contextual embedding of the query. The results in \FCR\ sorted based on their relevance scores and the top-$k$ ($k\leq3$) results are displayed to the user if their relevance scores exceed a predefined (empirically determined) cutoff. If the query language is a language other than English, the results are displayed after translating to the original language using the Language Processor module. If the API call does not return any result or the relevance scores of all the returned results are below the cutoff, the query is forwarded to the google-search module. A regional language query is forwarded to the Google search module in its original language. The flow-chart for the Fact Check module is shown in \Cref{fig:fake}.

\subsection{Google search articles}
\label{sec:google}

When a query is classified as GEN or the FAQ and Fact Check module does not produce any response to a query in FAQ or FAKE category respectively, the query is forwarded to the Google Search module that retrieves results by performing Google search using the google search library. The query is fed to the \texttt{search()} method of Google search library and the top two urls (already sorted based on their relevance to the query) are retrieved. For each of the two URLs obtained in the previous step, we scrape the webpage so that we can perform summarization using that scraped data. After scraping the url, the content is summarized using TextRank \cite{mihalceatarau2004textrank}, a graph-based summarization algorithm. To compute the edge weights of the graph used in TextRank, we use three different similarity metrics 
\begin{enumerate}
    \item Cosine similarity between sentences using TF-IDF model,
    \item Longest common substring between any two sentences, 
    \item BM25 / Okapi-BM25 \citep{robertson2009probabilistic}, a bag-of-words ranking function used for information retrieval tasks.
\end{enumerate}
The obtained summary and the url are displayed to the user. 
\begin{figure}
    \centering
    \includegraphics[width=\linewidth]{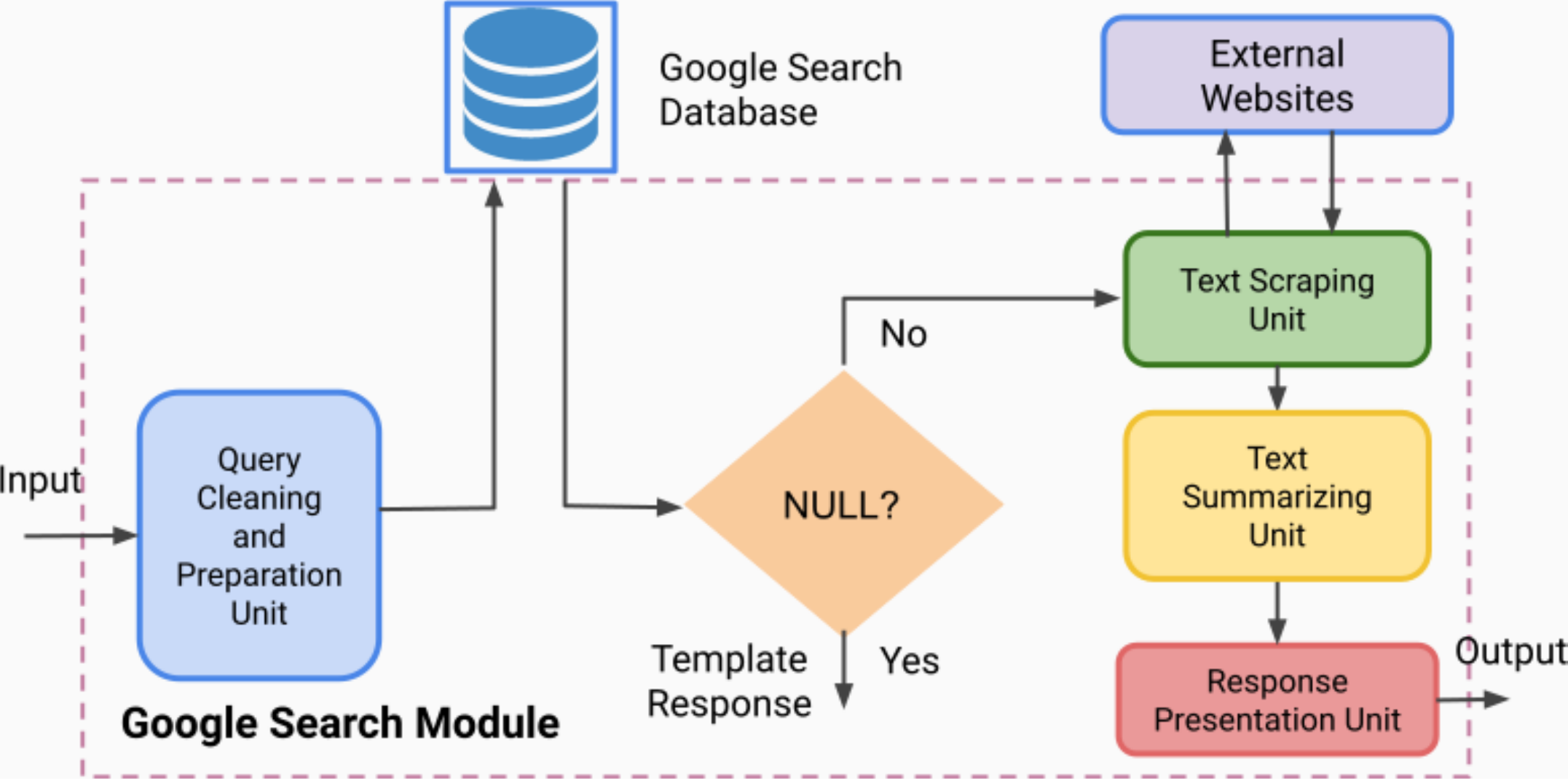}
    \caption{Consolidated architecture of Google search and summarizer layers.}
    \label{fig:google}
\end{figure}
If the original query is in a low-resource language and the scraped data of the webpage is in the same language, no translation is required. However, if the obtained webpages are in English, then the language processor (LP) module (see \Cref{fig:architecture}) translates the summary into the language of the original query before displaying it to the user.

\subsection{Extension to low-resource languages}
\label{sec:multilingual}

One distinguishing factor of \truthbot\ is that it is capable of responding in multiple languages. This is particularly helpful for the low-resource language users since the social media and instant messengers have significant user-base in these languages but technologies for various applications (including true information finding applications) are not well-developed for them. We use the language detection and translation service of Google translate\footnote{https://translate.google.com} in the QPU and language processor modules in \Cref{fig:architecture}. In QPU, if the detected language is different from English, then that information is stored, the query is translated into English and sent to the later modules. At the language processor module, the response of the bot is translated back into the original language of the query and passed through the conversational AI module to be displayed on user's screen.
\begin{figure}[t]
    \centering
    \includegraphics[width=0.85\linewidth]{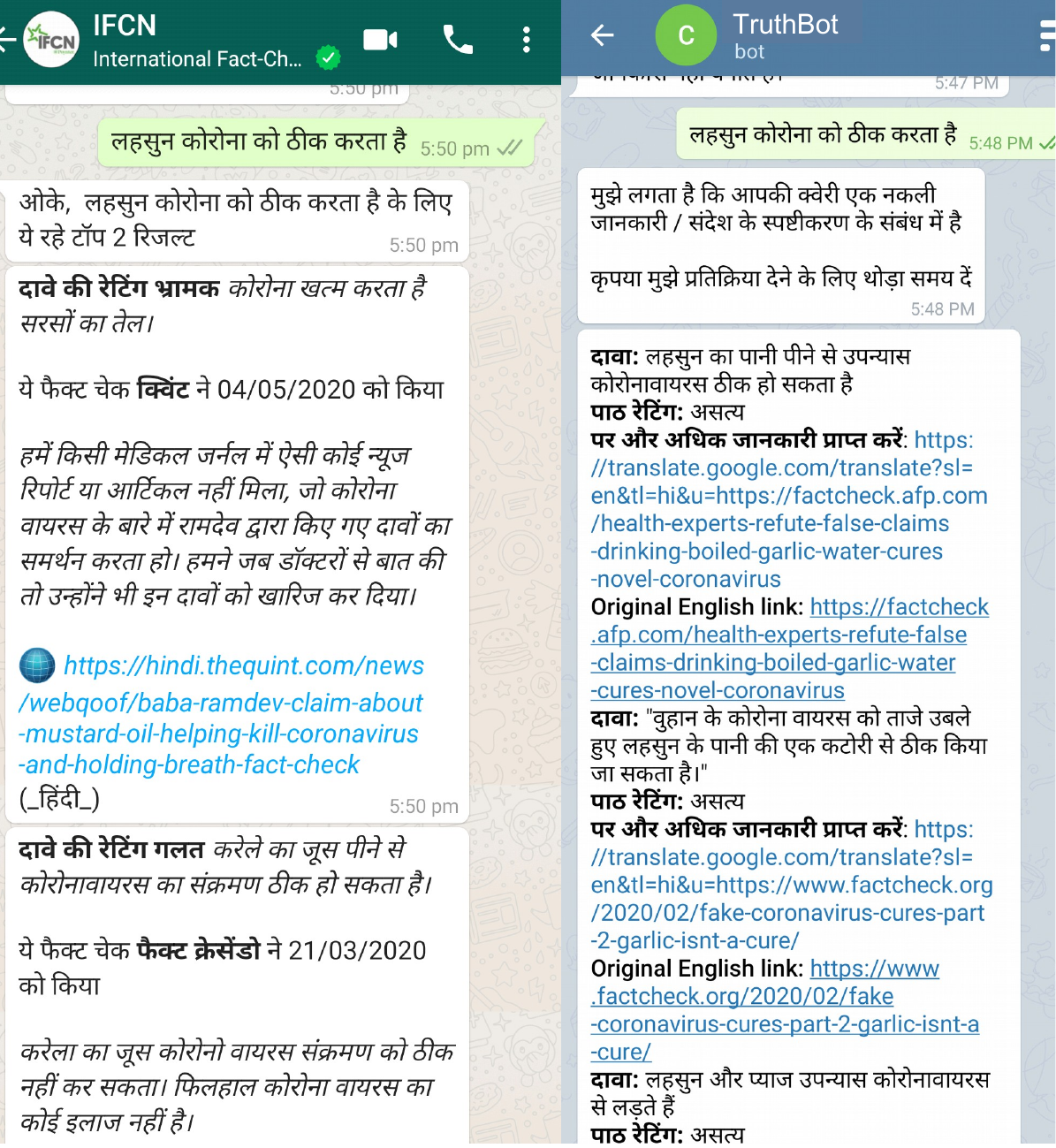}
    \caption{Response of IFCN chatbot (left) and \truthbot\ (right) in Hindi.}
    \label{fig:hindi}
\end{figure}
\Cref{fig:hindi} shows the responses of a typical query in {\em Hindi} in IFCN chatbot and in \truthbot.

\section{Experimental Results}
\label{sec:experiments}

The experiments are performed to validate the efficacy of \truthbot\ with its closest competitor, the IFCN chatbot. The comparison considers three metrics. First, we perform a quantitative comparison on the response accuracy, which compares the textual similarity of the query and the response provided by the two chatbots. Second, we run a user survey to capture their qualitative satisfaction with the responses of the two chatbots. Finally, we conduct a user survey about the user interface and the presentation of the curated information in these two chatbots.

\subsection{Evaluation dataset}
\label{ex_data}
To evaluate \truthbot, we curated a dataset consisting of $79$ queries. The queries belonged to four categories - FAQ, FAKE, GEN, and AREASTAT. The FAQ queries were manually curated from WHO and CDC which provide general information on COVID-19. For FAKE and GEN queries, we collected most recent (as of August 10, 2020) fake information and true news articles on COVID-19 from different media outlets (e.g., \url{https://www.indiatoday.in/}). Single sentence queries that conveyed the main message were curated from the collected articles. We curated the AREASTAT queries manually. Supplementary materials accompany the list of queries and the responses of the two chatbots. In the quantitative comparison of the response accuracy, we used the first three categories, since the textual similarity between AREASTAT queries and responses is not informative about accuracy.

\subsection{Evaluating response accuracy}
\begin{figure}[t!]
    \centering
    \includegraphics[width=\linewidth]{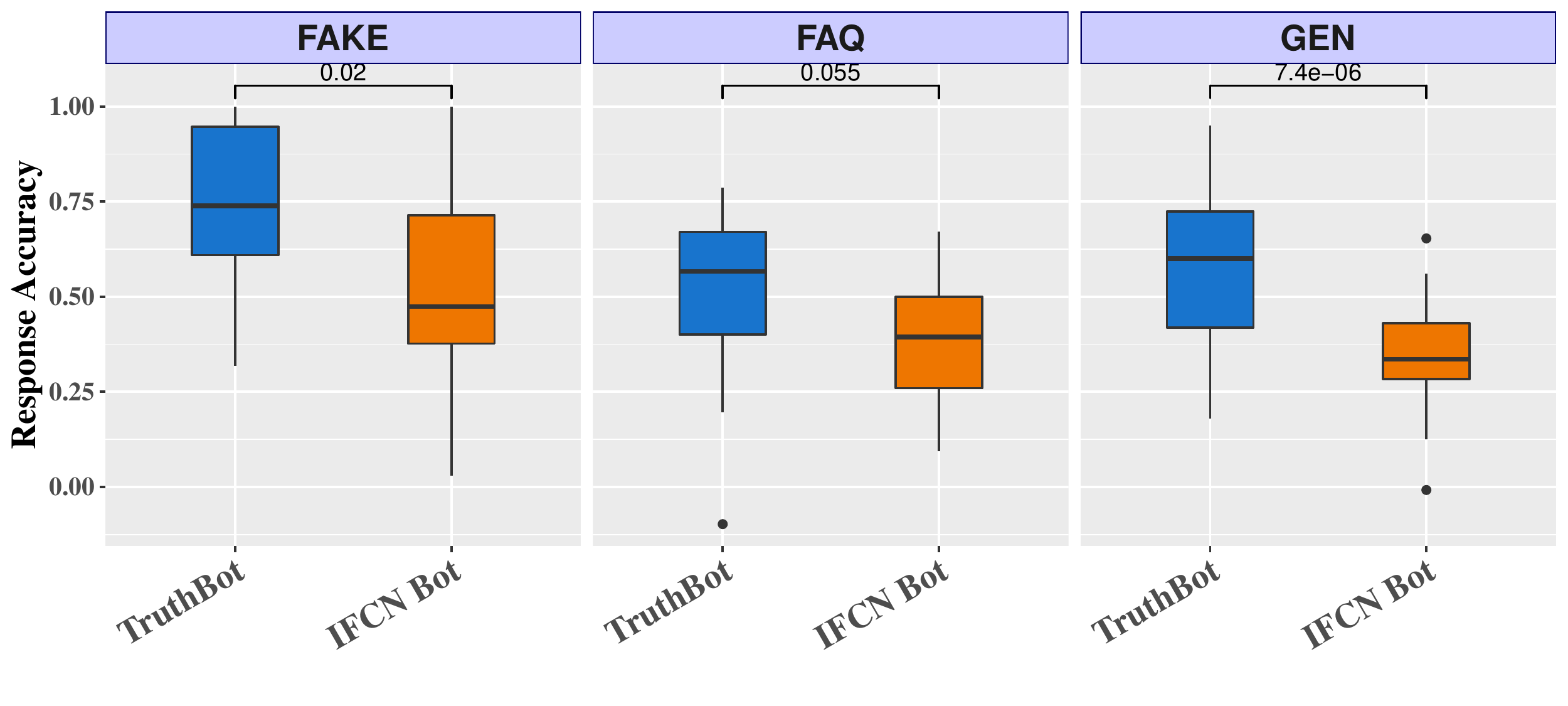}
    \caption{Comparison of Response Accuracy for FAKE, FAQ, and GEN queries. The numbers on the square brackets on the top of the plots denote the $p$-values of the $t$-test for statistical significance.}
    \label{fig:comp:RA}
\end{figure}

First, we wanted to evaluate the accuracy of \truthbot\ in retrieving the correct article available in the online or offline databases in response to the specific query. For this, we designed a metric called `response accuracy', which measures the similarity of the retrieved response against the actual query. We used cosine similarity between the main text of the retrieved response and the actual query for computing response accuracy. Results of \truthbot\ were compared against that of the chatbot by IFCN. For IFCN chatbot, the sentence following the phrase ``Claim rated false" or ``Claim rated misleading" was selected as the main text of the response. For \truthbot, for FAKE queries, the sentence followed by the phrase ``Claim" was selected as the main text, whereas for FAQ and GEN queries, the complete response was compared against that of the actual query. Fig. \ref{fig:comp:RA} compares the response accuracy of \truthbot\ against that of IFCN chatbot for different classes of queries. 
For FAKE queries, average response accuracy of \truthbot\ was $0.74$ as compared to $0.56$ for IFCN chatbot. Since IFCN is only meant for fake news alerting, the response that it produces for FAQ and GEN queries may not be meaningful. However, since the response accuracy measures the textual similarity between the response and the query, it also indicates whether relevant articles were retrieved by the chatbot. For FAQ and GEN queries, \truthbot's average response accuracy was $35.64\%$ and $68.2\%$ better than that of IFCN chatbot.

\subsection{Chatbot satisfaction survey}
\begin{figure}[t!]
    \centering
    \includegraphics[width=\linewidth]{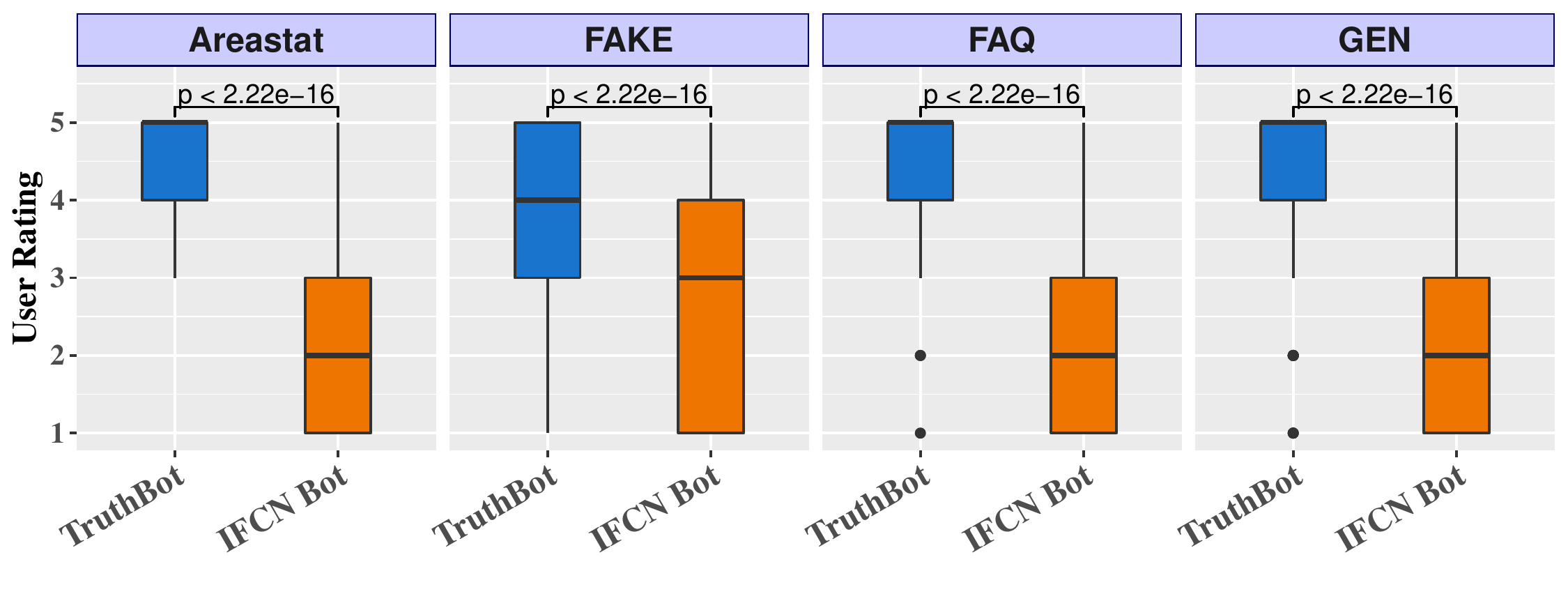}
    \caption{Comparison of user satisfaction while using \truthbot\ and IFCN chatbot based for AREASTAT, FAKE, FAQ and GEN queries. The results are based on a user survey. The numbers on the square brackets on the top of the plots denote the $p$-values of the $t$-test for statistical significance.}
    \label{fig:comp:US}
\end{figure}

We conducted a human-centric evaluation to determine how satisfied a user would be with the responses received from \truthbot\ as compared to IFCN chatbot. 
The subjects were chosen through an open call sent via email and 64 users participated in this survey. The age-group of the participants were: 10-30 years (80.3\%), 31-50 years (16.3\%), and above 50 years (3.4\%). Among them, 20.2\% were female and 79.2\% were male participants. Each user was assigned 10 queries (2 AREASTAT, 3 FAKE, 2 FAQ and 3 GEN) and the corresponding responses from \truthbot\ and IFCN chatbot. The queries were randomly chosen from the set of queries described in Section \ref{ex_data} so that we obtain multiple ratings (by multiple users) for a single query. Each user was instructed to rate the response of each bot based on how satisfied (1 for least and 5 for most) they were with the response obtained from a bot for each of their assigned queries. The result of the user survey is shown in \Cref{fig:comp:US}. 

For each of the query classes, \truthbot\ achieved much better user rating as compared to IFCN chatbot. For AREASTAT, FAQ and GEN, median user rating of \truthbot\ was 5 as compared to the median rating of 2 for IFCN chatbot. Even for FAKE queries (for which IFCN chatbot is specialized), \truthbot\ achieved better rating (median rating 4 as compared to 3 for IFCN chatbot).

\subsection{User interface study}
\begin{figure}[t!]
    \centering
    \includegraphics[width=\linewidth]{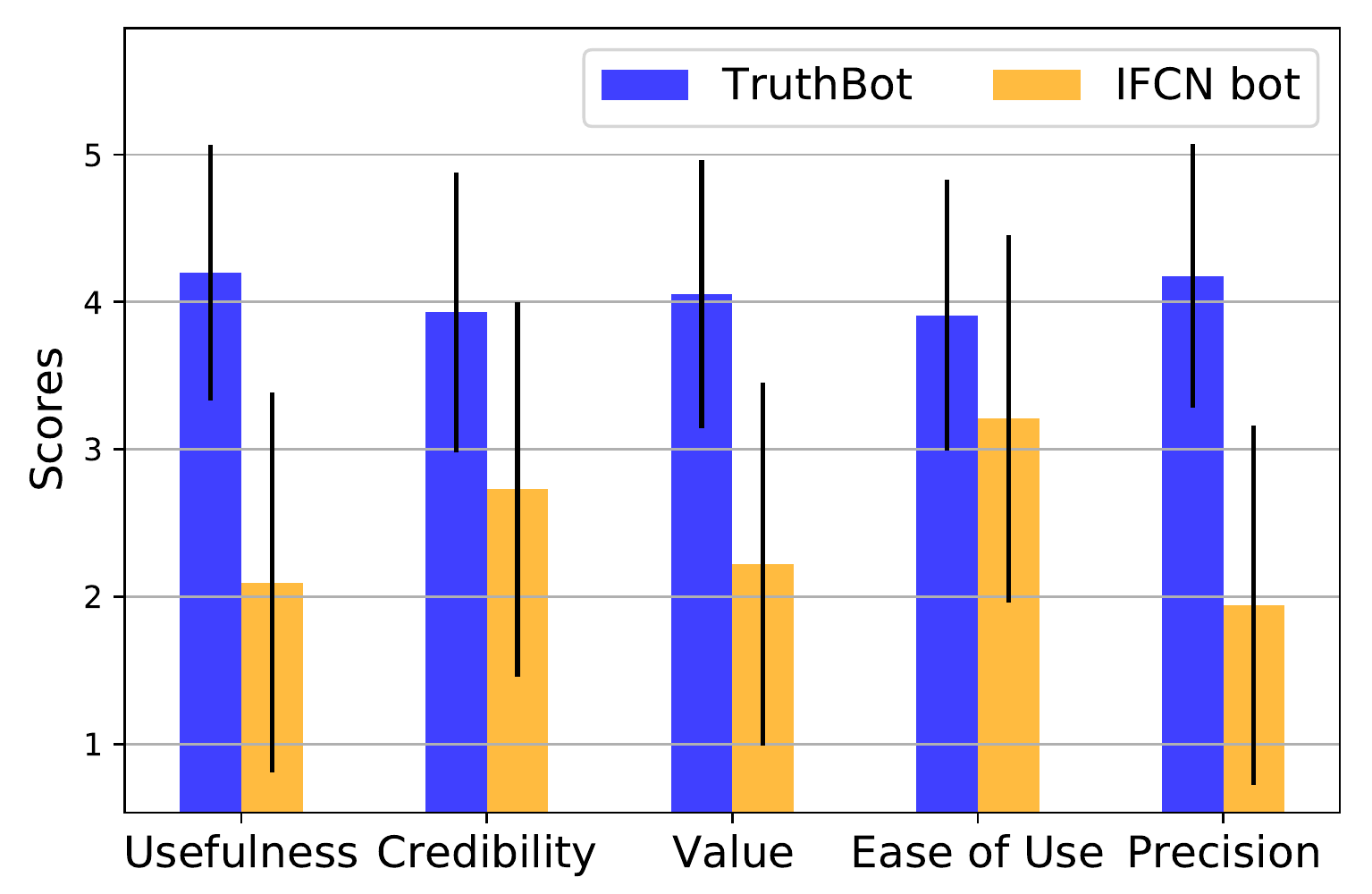}
    \caption{Comparison of the GUIs of the two chatbots along different metrics.}
    \label{fig:ui_comparison}
\end{figure}

Finally, we also compared \truthbot\ against the IFCN bot based on the feedback from the users regarding different aspects of the user interface - Usefulness, Ease of Use, Credibility, Precision, and Value~\citep{interaction-design.org2020}. We obtained responses from 51 users for this survey. In the survey, the users had to rate the bots for each of the above aspects. Possible ratings were from 1 (worst) to 5 (best). The result of the user survey is shown in \Cref{fig:ui_comparison}. 
For each of the above aspects, \truthbot\ obtained better ratings compared to that of IFCN bot. Specifically, for Usefulness and Precision, mean rating for \truthbot\ was 2 times better than that IFCN bot on an average. The $p$-values of the $t$-test for statistical significance on all the aspects were $<0.00001$.

\section{Summary and Future Work}

We presented a multilingual and multipurpose chatbot that provides almost all aspects of information on a topic a user can be interested in. Such chatbots will be a single-point destination for users during situations such as COVID-19. Performance evaluation and user survey (as exemplified for COVID-19) also demonstrate the superiority of the bot over the existing options (IFCN chatbot) in providing multi-faceted well-searched information on a specific topic. The bot can be customized for other topics too. A primary goal of this paper is to introduce and exhibit the concept of such holistic information dissemination through a chatbot and provide evaluations to show its efficacy. 

The chatbot still depends on a number of external databases. While this may not be completely avoidable, but in our future work, we plan to make some of these information to be cached in our local databases to reduce the dependencies on such external factors, which helps both in the response latency and reliability. We also plan to expand the scope of the chatbot to image and multimedia messages, which are often sources of misinformation.

\bibliographystyle{plainnat}
\bibliography{abb,swaprava,ultimate,references,references_robots,master}

\end{document}